\documentclass[twocolumn,showpacs,prd]{revtex4}
\usepackage{epsfig}

\begin{document}

\title{Lepton asymmetry and primordial nucleosynthesis
in the era of precision cosmology}

\author{Pasquale D.~Serpico and Georg~G.~Raffelt}
\affiliation{Max-Planck-Institut f\"ur
Physik (Werner-Heisenberg-Institut), F\"ohringer Ring 6, 80805
M\"unchen, Germany}

\date{\today}

\begin{abstract}
 We calculate and display the primordial light-element abundances as a function 
 of a neutrino degeneracy parameter $\xi$ common to all
 flavors.  It is the only unknown parameter characterizing the thermal medium at 
 the primordial nucleosynthesis epoch.  The observed
 primordial helium abundance $Y_{\rm p}$ is the most sensitive cosmic 
 ``leptometer.''  Adopting the conservative $Y_{\rm p}$ error analysis of Olive 
 and Skillman implies $-0.04\alt \xi \alt 0.07$ whereas the errors stated by 
 Izotov and Thuan imply $\xi=0.0245\pm 0.0092$
 ($1\,\sigma$).  Improved determinations of the baryon abundance have no 
 significant impact on this situation.  A determination of $Y_{\rm p}$ that 
 reliably distinguishes between a vanishing or nonvanishing $\xi$ is a crucial 
 test of the cosmological standard assumption that sphaleron effects equilibrate 
 the cosmic lepton and baryon asymmetries.
\end{abstract}

\pacs{26.35.+c,%Big Bang nucleosynthesis
 14.60.Lm,     %Ordinary neutrinos 
 98.80.Es      %Observational cosmology 
\hfill Preprint MPP-2005-54}

\maketitle

%%%%%%%%%%%%%%%%%%%%%%%%%%%%%%%%%%%%%%%%%%%%%%%%%%%%%%%%%%%%%%%%%%%%%%
\section{Introduction}
%%%%%%%%%%%%%%%%%%%%%%%%%%%%%%%%%%%%%%%%%%%%%%%%%%%%%%%%%%%%%%%%%%%%%%

The successful calculation of the primordial light-element abundances
in the framework of the standard theory of big-bang nucleosynthesis
(BBN)~\cite{Wagoner:1966pv, Olive:1999ij, Burles:2000zk,
  Eidelman:2004wy} is among the main pillars supporting our modern
understanding of the universe. Apart from the cross sections that are
needed for the network of nuclear reactions, these predictions depend
on the baryon-to-photon ratio $\eta_B$, on the radiation density at
the BBN epoch that is traditionally parametrized by the equivalent
number of neutrino flavors $N_{\rm eff}$, on the electron-neutrino
degeneracy parameter $\xi=\mu_{\nu_e}/T_{\nu_e}$, and on the
degeneracy parameters of the other neutrino flavors.  Originally all
of these parameters had to be fitted such as to reproduce the observed
light-element abundances.

Meanwhile the number of neutrino flavors has been fixed by the $Z^0$
decay width~\cite{Eidelman:2004wy}. More recently, the baryon
abundance has been determined by the cosmic microwave background (CMB)
anisotropies to be~\cite{Spergel:2003cb}
\begin{equation}\label{eq:baryons}
\eta_B=\frac{n_{B}-n_{\bar B}}{n_\gamma}
=6.14\times10^{-10} (1.00\pm0.04)\,.
\end{equation}
Finally, the measured neutrino mixing parameters imply that neutrinos
reach approximate chemical equilibrium before the BBN epoch so that
all neutrino chemical potentials can be taken to be equal, i.e.\ they
are all characterized by the same degeneracy parameter $\xi$ that
applies to $\nu_e$ \cite{Dolgov:2002ab,Wong:2002fa,Abazajian:2002qx}.

Of course, one may still assume new low-energy physics such as
sizeable active-sterile neutrino mixings, the existence of new
particles such as axions that could contribute to the primordial
radiation density, a time-varying Newton constant, and a plethora of
other non-standard options that would affect BBN~\cite{Sarkar:1995dd}.
Barring these more exotic possibilities, the single remaining BBN
parameter that is not otherwise fixed is $\xi$. In the light of these
developments we re-examine BBN under the assumption that
$\xi$ is the only free input parameter.

%%%%%%%%%%%%%%%%%%%%%%%%%%%%%%%%%%%%%%%%%%%%%%%%%%%%%%%%%%%%%%%%%%%%%%
\section{BBN with a lepton asymmetry}
%%%%%%%%%%%%%%%%%%%%%%%%%%%%%%%%%%%%%%%%%%%%%%%%%%%%%%%%%%%%%%%%%%%%%%

\begin{figure}%[!htb]
\epsfig{file=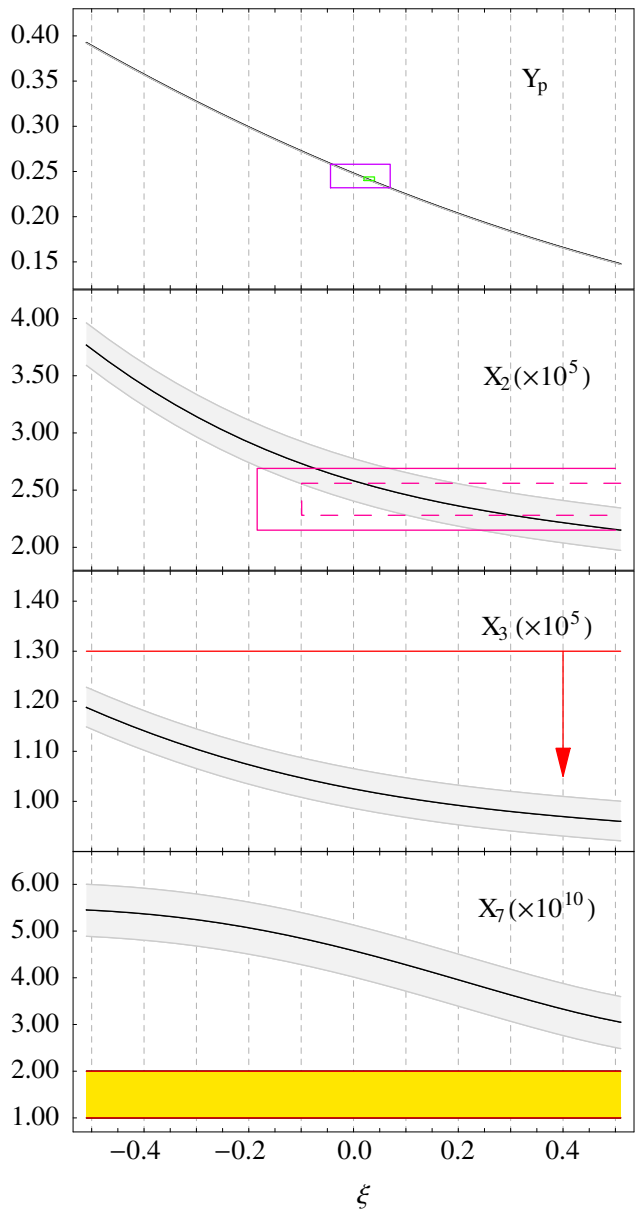,width=\columnwidth}
\caption{Light-element abundances as a function of the neutrino
  degeneracy parameter. The top panel shows the primordial $^4{\rm
    He}$ mass fraction $Y_{\rm p}$ whereas the other panels show the
  $^2{\rm H}$, $^3{\rm He}$, and $^7{\rm Li}$ number fractions
  relative to hydrogen. The gray $1\,\sigma$ error bands include the
  uncertainty of the WMAP determination of the baryon abundances well
  as the uncertainties from the nuclear cross
  sections~\cite{Serpico:2004gx}. We also show the observed abundances
  as described in the text. \label{figure1}}
\end{figure}

\begin{figure}%[!htb]
\epsfig{file=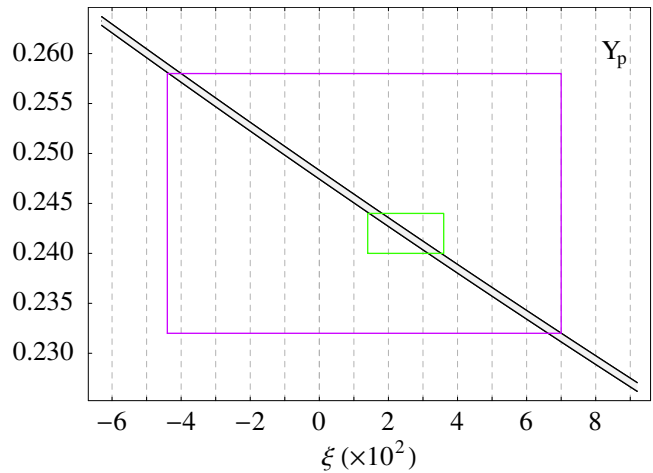,width=\columnwidth}
\caption{A blowup of the first panel of Fig.~\ref{figure1}, showing
  the $1\,\sigma$ range for the primordial helium mass fraction. The
  small box represents the $1\,\sigma$ observational error quoted by
  Izotov and Thuan~\cite{Izotov:2003xn} whereas the large box
  represent the maximum systematic error quoted by Olive and
  Skillman~\cite{Olive:2004kq}.
\label{figure2}}
\end{figure}

Over the years, BBN with a lepton asymmetry has been studied by many
authors~\cite{Wagoner:1966pv, Freese:1982ci, Kang:1991xa,
Esposito:2000hi, Esposito:2000hh, Barger:2003rt, Cuoco:2003cu,Kneller:2004jz}. 
To achieve approximate agreement between the observed and predicted
light-element abundances a possible lepton asymmetry must be small so
that one may well constrain a new analysis, say, to the range
$|\xi|<0.5$.  For such small $\xi$ values the most important impact on
BBN is a shift of the beta equilibrium between protons and neutrons.
Subleading effects include a modification of the radiation density,
\begin{equation}\label{eq:deltanu}
\Delta N_{\rm eff}=
3\bigg[\frac{30}{7}\bigg(\frac{\xi}{\pi}\bigg)^2
+\frac{15}{7}\bigg(\frac{\xi}{\pi}\bigg)^4\bigg]\,.
\end{equation}
Moreover, the neutrino decoupling temperature is higher than in the
standard case~\cite{Freese:1982ci, Kang:1991xa} so that in principle
one could get a non-standard $T_\nu(T)$ evolution, but such effects
are completely negligible for our case.  A non-zero $\xi$ slightly
modifies the partial neutrino reheating following the $e^{+}e^{-}$
annihilation~\cite{Esposito:2000hi}, again a completely negligible
effect for our $\xi$ range of interest.

To include the modified $n\leftrightarrow p$ weak rates we
adopt a perturbative approach for the modified neutrino distribution
functions, similar to the one adopted by us in a different
context~\cite{Serpico:2004nm}.  Therefore, the additional terms for
the rates can be obtained by integrating the factors
\begin{equation}
\delta f_\nu (E,\xi)=
\sum_{n=1}^\infty
\frac{\partial f_\nu(E,\xi)}{\partial \xi}\bigg{|}_{\xi=0}
\frac{\xi^n}{n!}.
\end{equation}
Corrections up to fourth order were included into the code of
Ref.~\cite{Serpico:2004gx}, thus reducing the error of the truncation
on the nuclide yields to the fourth significant digit for $|\xi|\le
0.1$, and to the third for $|\xi|\le 0.5$.  This also sets the level
of the required fitting accuracy of the auxiliary functions.  The
corrections were calculated as relative changes with respect to the
$\xi=0$ Born rates so that in the $\xi\to 0$ limit we recover the
standard $n\leftrightarrow p$ rates with finite mass, QED radiative
and thermodynamic corrections, and partial neutrino
reheating~\cite{Serpico:2004gx}.  For $\xi\not=0$ these sub-leading
effects are rescaled without introducing significant errors.  We have
checked the reliability of our approach by comparing with the code of
Ref.~\cite{Esposito:2000hh} where a fully numerical and
non-perturbative method is used.

In Fig.~\ref{figure1} we show our predictions for the primordial
light-element abundances as a function of the neutrino degeneracy
parameter $\xi$, taken to be equal for all flavors.  The gray band is
the $1\,\sigma$ predicted range.  We used the nuclear reactions and
uncertainties adopted in Ref.~\cite{Serpico:2004gx}.  However, except
for lithium the uncertainty is dominated by the error of
the Wilkinson microwave anisotropy probe (WMAP)-implied~\cite{Spergel:2003cb} 
baryon abundance of Eq.~(\ref{eq:baryons}).  In Fig.~\ref{figure2} we show a 
blowup of the top panel, i.e.~the predicted helium abundance.

We also show observations of the primordial abundances.  Beginning
with $^7\rm{Li}$ (bottom panel of Fig.~\ref{figure1}) we show the
allowed range $\log_{10}(X_7)+12=2.0$--2.3, where the uncertainty on
$T_{\rm eff}$ has been included~\cite{Lambert:2004kn}.  In the second
to last panel we show Bania et~al.'s possible upper limit on the
$^3$He abundance of $1.3\times 10^{-5}$~\cite{Bania2002} (at
$1\,\sigma$).  The status and the interpretation of these observations
is quite controversial, especially for $^7{\rm Li}$; see
Ref.~\cite{Serpico:2004gx} for further discussions and references.  It
is evident, however, that the $^3{\rm He}$ upper limit is compatible
with any realistic $\xi$ value whereas $^7{\rm Li}$ is always
incompatible.  Rather than dismissing standard BBN as the correct
theory for the light-element synthesis we follow the usual practice of
dismissing the $^7{\rm Li}$ observations as too uncertain to be
useful. Putting this more positively, it is remarkable how close the
$^7{\rm Li}$ abundances observed on the Spite plateau in metal poor
stars come to the BBN prediction.

Turning to deuterium (second panel in Fig.~\ref{figure1}), we note
that a new detection was reported, giving $X_2\equiv {}^2{\rm H}/{\rm
H}=1.60_{-0.30}^{+0.25}\times 10^{-5}$ \cite{Crighton:2004aj},
significantly lower than the previous value towards Q1243+3047 of
$2.42_{-0.25}^{+0.35}\times 10^{-5}$ \cite{Kirkman:2003uv}.  By
adopting the conservative approach to symmetrize the errors to the
higher value, and the wider allowed range (1.57--$4.00)\times 10^{-5}$
for the most uncertain determination towards Q0347-3819, one gets
$2.42\pm 0.27$ (outer error box in Fig.~\ref{figure1}) for the average
value of the seven positive detections so far obtained in quasar Lyman
absorption systems.  As previously noted~\cite{Kirkman:2003uv}, since
$\chi^2$ per degree of freedom is approximately 4.4, even with these
generous error estimates, underestimated systematics may be present.

Still, the observed primordial deuterium abundance looks perfectly
compatible with the standard BBN predictions in the absence of a
lepton asymmetry. Neglecting possible systematics one finds a lower
limit $\xi>-0.2$ at $1\,\sigma$, but no useful upper limit obtains.
This remains true even if the precision of the deuterium observations
improve by a factor of two (inner box in Fig.~\ref{figure1}).

From Figs.~\ref{figure1} and~\ref{figure2} it is obvious that $^4{\rm
He}$ is the best probe of $\xi$.  As the dependence on the baryon
abundance $\eta_B$ is only logarithmic, the new WMAP data do not
significantly improve the constraint.  Therefore, the observed range
for $Y_{\rm p}$ is the only truly significant measure for a possible
lepton asymmetry.

Recently new data were published by Izotov and
Thuan~\cite{Izotov:2003xn}, giving $Y_{\rm p}=0.2421\pm 0.0021$ (inner
error box in Figs.~\ref{figure1} and~\ref{figure2}).  Slightly higher
values arise for differently chosen samples and/or for other
metallicity regressions.  For the purpose of illustration we have
performed a likelihood analysis of the BBN prediction for $Y_{\rm
  p}(\eta_B,\xi)$ using the $\eta_B$ prior of Eq.~(\ref{eq:baryons})
and Izotov and Thuan's $Y_{\rm p}$ with their quoted statistical
error.  The result suggests a positive value $\xi=0.0245\pm0.0092$
($1\,\sigma$), i.e.~a hint for $\xi\not=0$ at $2.7\,\sigma$.  This
exercise illustrates the possible sensitivity that present
determinations of $^4$He could reach if the systematic uncertainties
were fully understood.

In an independent analysis Olive and Skillman~\cite{Olive:2004kq} find
$Y_{\rm p}=0.249\pm 0.009$.  In an attempt to quantify systematic
effects, they suggest as the most conservative range $0.232\leq Y_{\rm
  p} \leq 0.258$ (outer box in Figs.~\ref{figure1} and~\ref{figure2}),
corresponding to $-0.044\alt\xi\alt0.070$ for the $1\,\sigma$ range of
$\eta_B$ or $-0.046\alt\xi\alt0.072$ for the $2\,\sigma$ range.

To compare $\xi$ with $\eta_B$ on the same footing we note that the
cosmic abundance of a lepton flavor $\alpha$ is given in terms of the
corresponding degeneracy parameter as
\begin{equation}
\eta_\alpha=\frac{n_{\alpha}-n_{\bar \alpha}}{n_\gamma}
=\frac{1}{12\zeta(3)}\left(\frac{T_{\alpha}}{T_\gamma}\right)^3
\left(\pi^2\xi_\alpha+\xi_\alpha^3\right)\,.
\end{equation}
Charge neutrality implies that $\eta_e$ is of the same order as
$\eta_B$. A possible large lepton asymmetry thus resides in neutrinos,
\begin{equation}
\eta_L\approx\sum_\nu\eta_\nu\approx3\times0.249\,\xi\,,
\end{equation}
where $(T_\nu/T_\gamma)^3=4/11$ was assumed.  Note that this factor is
missing in Eq.~(1) of Ref.~\cite{Barger:2003rt} and Eq.~(10) of 
Ref.~\cite{Kneller:2004jz}, that limits their validity to temperatures
larger than the electron mass.
%%%%%%%%%%%%%%%%%%%%%%%%%%%%%%%%%%%%%%%%%%%%%%%%%%%%%%%%%%%%%%%%%%%%%%
\section{Discussion and Summary}
%%%%%%%%%%%%%%%%%%%%%%%%%%%%%%%%%%%%%%%%%%%%%%%%%%%%%%%%%%%%%%%%%%%%%%

We have argued that after the experimental determination of the
neutrino mixing parameters and after the WMAP determination of the
baryon abundance, of all parameters that characterize the cosmic
thermal heat bath at the BBN epoch only the lepton asymmetry, i.e.\ a
neutrino degeneracy parameter $\xi$ common to all flavors, remains
undetermined and has to be fixed by the observed light-element
abundances. For the first time we have produced a meaningful plot of
the predicted light-element abundances as a function of $\xi$.  One
motivation for writing this short note was to present for the first
time this plot as an alternative to the still common but outdated
depiction of the light-element abundances as a function of $\eta_B$.

The errors of the predictions in our plot are typically 
dominated by the WMAP determination of the baryon fraction $\eta_B$ 
and thus will significantly improve with forthcoming CMB missions.

The deuterium abundance is very sensitive to $\eta_B$ so that
deuterium is the best baryometer of all the light elements. On the
other hand, this element does not respond much to a non-zero lepton
asymmetry. The opposite applies to helium which is by far the most
sensitive leptometer whereas it is virtually insensitive to future
improvements of $\eta_B$ determinations.

The usual attitude towards the possibility of a large cosmic lepton
asymmetry is that sphaleron effects before electroweak symmetry
breaking equilibrate the cosmic lepton and baryon asymmetries to
within a factor of order unity. In the standard model one finds that
$\eta_L=-\frac{51}{28} \eta_B$. From this perspective, BBN is a
parameter-free theory and the cosmic helium abundance is fixed by the
$\xi=0$ value shown in Fig.~\ref{figure2}, i.e.\ $Y_{\rm p}=0.2479\pm
0.0005$ (1 $\sigma$). In this scenario it remains to understand the systematic
errors in the spectroscopic $Y_{\rm p}$ determinations.

However, we think that this argument should be turned around.
Sphaleron effects are a crucial ingredient in most baryogenesis
scenarios~\cite{Dine:2003ax}, including
leptogenesis~\cite{Buchmuller:2005eh}. No experimental evidence for or
against these effects exists. Therefore, a possible indication for a
nonzero $\xi$ by the observed helium abundance presents an approach
for a possible falsification of the usual assumptions about
baryon-lepton equilibration.

In principle, other cosmological probes of a non-zero $\xi$ exist in
the form of the CMB anisotropies and large-scale structure (LSS) power
spectra~\cite{Kinney:1999pd,Lesgourgues:1999wu}.  The former
essentially feels a boost in the amplitude of the first peak and a
shift towards higher multipoles, effects that are degenerate with
$N_{\rm eff}>3$. The latter suffers a suppression very similar to the
effect of a non-zero $m_\nu$.  However, the current CMB and LSS data
are only sensitive to $\xi= {\cal O}(1)$~\cite{Hannestad:2003xv}.

A step further may be provided by future CMB limits on $\Delta N_{\rm eff}$,
possibly helped by accurate polarization maps~\cite{Lopez:1998aq}.
Assuming Kinney and Riotto's estimate~\cite{Kinney:1999pd} of the
capability of PLANCK to detect one degeneracy parameter of order
$0.5$, we can easily translate it into a marginal sensitivity of
$\xi\simeq 0.3$ that is common to all three flavors. However, unless
one is able to break the degeneracies with other cosmological
parameters, such a level of accuracy is too optimistic.

To reach a sensitivity comparable with BBN, one has to wait for more
ambitious methods studying the gravitational lensing distortions on
both the temperature and polarization maps of the
CMB~\cite{Kaplinghat:2003bh}.  Emphasis is usually put on the
sensitivity to neutrino masses, but these measurements are useful to
constrain the relic neutrino asymmetry as well, especially if
complementary information would be provided by improved LSS maps and
direct experimental data on the neutrino masses.  Incidentally, such
an accurate satellite mission would be sensitive to $Y_{\rm p}$ at the
recombination epoch at a level of $\delta Y_{\rm p}\alt
0.005$~\cite{Kaplinghat:2003bh}, thus comparable or better than
present astrophysical spectroscopic determinations, with a positive
impact on the BBN constraint on $\xi$.  In any event, BBN is the only
probe sensitive to the sign of $\xi$.

In summary, now that the cosmic baryon abundance has been extremely
well determined by CMB observations and now that much about neutrino
properties has been learnt by experiments, the role of BBN as a
baryometer has shifted to that of the best available cosmic
leptometer.  Therefore, a more reliable $Y_{\rm p}$ determination is
of much greater fundamental interest than the next round of more
precise CMB baryon determinations.

%%%%%%%%%%%%%%%%%%%%%%%%%%%%%%%%%%%%%%%%%%%%%%%%%%%%%%%%%%%%%%%%%%%%%%
\section*{ACKNOWLEDGMENTS} %%%%%%%%%%%%%%%%%%%%%%%%%%%%%%%%%%%%%%%%%%%
%%%%%%%%%%%%%%%%%%%%%%%%%%%%%%%%%%%%%%%%%%%%%%%%%%%%%%%%%%%%%%%%%%%%%%

We acknowledge partial support by the Deutsche Forschungsgemeinschaft
under Grant No.~SFB-375 and by the European Union under the Ilias
project, contract No.~RII3-CT-2004-506222.

%%%%%%%%%%%%%%%%%%%%%%%%%%%%%%%%%%%%%%%%%%%%%%%%%%%%%%%%%%%%%%%%%%%%%%
%% References %%%%%%%%%%%%%%%%%%%%%%%%%%%%%%%%%%%%%%%%%%%%%%%%%%%%%%%%
%%%%%%%%%%%%%%%%%%%%%%%%%%%%%%%%%%%%%%%%%%%%%%%%%%%%%%%%%%%%%%%%%%%%%%

%%%%%%%%%%%%%%%%%%%%%%%%%%%%%%%%%%%%%%%%%%%%%%%%%%%%%%%%%%%%%%%%%%%%%%
\end{document}